\documentclass[a4paper, amsfonts, amssymb, amsmath, reprint]{revtex4-2}

\usepackage[english]{babel}
\usepackage[utf8]{inputenc}
\usepackage[pdftex, pdftitle={Article}, pdfauthor={Author}]{hyperref} 
\usepackage{layouts}

\usepackage[T1]{fontenc}
\usepackage{graphicx,import}

\begin{document}


\title{Stability of Optically Injected Two-State Quantum-Dot Lasers}

\author{Stefan Meinecke}
    \affiliation{Institut für Theoretische Physik, Technische Universität Berlin, Hardenbergstr. 36, 10623 Berlin, Germany}

\author{Benjamin Lingnau}
    \affiliation{Institut für Theoretische Physik, Technische Universität Berlin, Hardenbergstr. 36, 10623 Berlin, Germany}

\author{André Röhm}
    \affiliation{Institut für Theoretische Physik, Technische Universität Berlin, Hardenbergstr. 36, 10623 Berlin, Germany}

\author{Kathy Lüdge}
    \affiliation{Institut für Theoretische Physik, Technische Universität Berlin, Hardenbergstr. 36, 10623 Berlin, Germany}

\begin{abstract}
  Simultaneous two-state lasing is a unique property of semiconductor quantum-dot (QD) lasers. This not only changes steady-state characteristics of the laser device but also its dynamic response to perturbations. In this paper we investigate the dynamic stability of QD lasers in an external optical injection setup. Compared to conventional single-state laser devices, we find a strong suppression of dynamical instabilities in two-state lasers. Furthermore, depending on the frequency and intensity of the injected light, pronounced areas of bistability between both lasing frequencies appear, which can be employed for fast optical switching in all-optical photonic computing applications. These results emphasize the suitability of QD semiconductor lasers in future integrated optoelectronic systems where a high level of stability is required.
\end{abstract}
% \shortabstract

{\onecolumngrid
This is the pre-peer reviewed version of the following article: Stability of Optically Injected Two-State Quantum-Dot Lasers, S. Meinecke, B. Lingnau, A. Röhm, K. Lüdge, ANNALEN DER PHYSIK 2017, 529, 1600279, which has been published in final form at \url{https://doi.org/10.1002/andp.201600279}. This article may be used for non-commercial purposes in accordance with Wiley Terms and Conditions for Use of Self-Archived Versions.
}
\clearpage

\maketitle
%%% Use this if the article text won't start with a \section:
% \noindent

\section{Introduction}

Quantum-dot (QD) based semiconductor lasers have a variety of beneficial properties, e.g. low threshold current, high differential gain, small line width and weak temperature sensitivity, which make them appealing for innovative applications in optical communication networks \cite{CHO13a,LUE11a,BIM99}. Quantum dots are most commonly self-assembled nanometer-sized semiconductor structures embedded in a surrounding quantum well. They provide confinement in all spatial dimensions, which results in localized electronic states and a discrete density of states. Relatively slow charge-carrier dynamics between the localized states \cite{BIM08a} then allow for simultaneous lasing from different electronic states at distinct wavelengths. Lasing has been experimentally demonstrated for emission at two \cite{MAR03a,SUG05b} and even three \cite{ZHA10a} wavelengths. Simultaneous lasing from the QD ground (GS) and first excited (ES) state is called \textit{two-state lasing}. Additionally, some devices can exhibit a decreasing emission from the GS wavelength with increasing pump current, resulting in the complete turn-off of the GS emission (quenching) and a switching to higher-energy emission. Since both wavelengths also have different threshold pump currents, there are typically three distinct regimes of operation: \textit{ground-state lasing}, \textit{two-state lasing}, and \textit{excited-state lasing}. The modeling of \textit{two-state lasing} has been the topic of previous research\cite{VIK05,JI10,GIO12,KOR13,ROE15a}. Applications of \textit{two-state lasing} effects are discussed for free-running all-optical gating \cite{VIK16}, feedback induced switching \cite{VIR14a}, anti-phase excitability \cite{KEL16} and self-mixing velocimetry \cite{GIO14}.

Our optical injection setup considers an external laser, that drives the QD laser at a slightly detunend frequency. This method has been widely studied for quantum well and single-color QD lasers and can be used to stabilize the emitted frequency \cite{TAR95a,SIM97a, WIE05, ERN10a}. We want to concentrate on the effects that arise due to the ability of the QD laser to exhibit \textit{two-state lasing}, specifically the role of the resulting higher dimensional phase-space, which, surprisingly, improves the dynamical stability. An increased stability is also theoretically predicted when the microscopic polarization introduces an additional degree of freedom into the system dynamics \cite{WIE16}, which possibly suggests a counter-intuitive but general trend towards higher stability with increasing number of degrees of freedom in semiconductor lasers.

We will discuss the stability boundaries by exploring the parameter space of the injected frequency and intensity. Our aim is to analyse the influence of driving current and amplitude-phase coupling upon the bifurcation diagram, especially on the optical bistability between the two different lasing frequencies. A bistability between two different locked solutions already occurs in single-state QD laser with optical injection\cite{ERN10a}, however, we focus on the optical bistability between two wavelengths that was recently demonstrated experimentally for QD lasers in an all-optical switching setup \cite{TYK14,TYK16a}. Previously, stability boundaries have been derived analytically in the context of two-polarization-mode lasing with optical injection \cite{FRI15} and elaborate bifurcation analysis was performed for optically injected Fabry-P\'{e}rot cavities that support two longitudinal modes \cite{OSH14,OSB12,OSB09,OSB09a}.  Yet, in the case of two-state QD lasers, the underlying dynamics are different as the two wavelengths emerge from different localized levels with a large energy separation of about 70meV. Thus, they do not directly compete for the same gain medium which on the one hand requires a modeling with more dynamic variables and on the other hand gives rise to new dynamics as for example ground state quenching \cite{ROE15a}. For our investigations, we consider a 1mm long edge-emitting laser based on InAs/InGaAs QDs, which consists of 15 QD layers, each $2\mu m$ wide and $4nm$ tall. The GS emits at $\hbar \omega^{GS}= 0.952eV$ and the ES at $\hbar \omega^{ES} = 1.022eV$ ($1300$nm and $1210$nm, respectively.).

 Theoretical results for two-state lasers with optical injection suggest the inhomogeneous broadening of QD states to be crucial for the observation of the hysteresis \cite{TYK16a}. The authors also emphasize the need for a model that goes beyond the $\alpha$-factor approximation to describe the amplitude-phase coupling of the gain medium. In our results, however, we show that the bistability is directly caused by the GS amplitude-phase coupling and that the usual linear $\alpha$-factor approximation can be sufficient, if the driving current and the strength of the amplitude-phase coupling is tuned correctly.

\begin{figure}
  \centering
  \includegraphics[width=0.7\columnwidth]{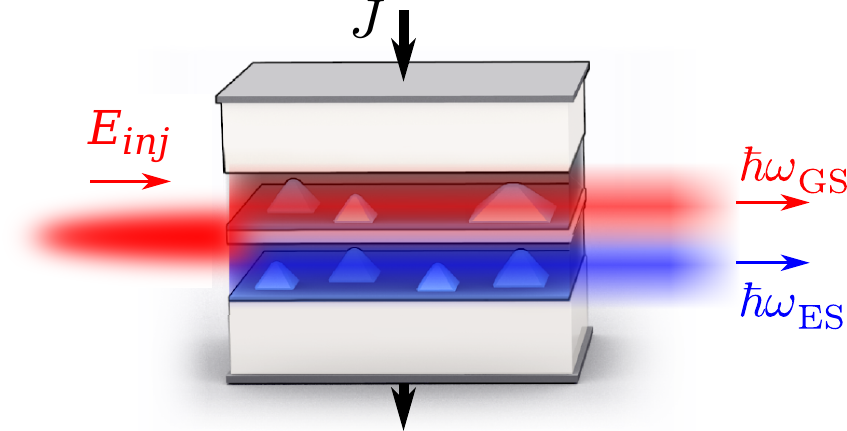}
  \caption{Sketch of the optically injected two-state QD laser. The injection $E_{inj}$ is at the ground state wavelength, while the resulting dynamics at the ground state energy ($\hbar \omega^{GS}$) and excited state energy ($\hbar \omega^{ES}$) are investigated.}\label{fig:setup}
\end{figure}

\section{Model \& Equations}

The model used within this paper is a rate-equation model that is based upon semiconductor-Bloch-equations with microscopically calculated charge-carrier scattering rates and builds upon previous work \cite{MAJ11,LIN14,ROE15a}. By comparison with the full semiconductor-Bloch-equation approach we extracted the frequency shift $\delta \omega$ of the GS electric field induced by changes in the entire carrier distribution \cite{LIN14}, whereas the conventional $\alpha$-factor approach only takes into account the influence of the active carrier population on the refractive index. Thus, the phase dynamics of the field inside the QD-laser cavity, i.e., the instantaneous frequency shift, is decoupled from the gain dynamics in our model, which makes the description more accurate than using a linear $\alpha$-factor \cite{LIN12}. The equation for the ground and excited state field amplitudes ($E^{GS}, E^{ES}$) are given in Eq.\,\eqref{Egs}-\eqref{Ees}.

\begin{align}
  \dot{E}^{GS} = & \left[ g_{GS} (\rho_{e,\textit{act}}^{GS} + \rho_{h,\textit{act}}^{GS} - 1) -i \gamma \delta \omega - \kappa^{GS} \right] E^{GS} \nonumber \\
		 & + K \kappa^{GS} E_0 e^{-i 2 \pi \Delta \nu_{inj} t} \label{Egs}\\
   \dot{E}^{ES} = & \left[ g_{ES} (\rho_{e,\textit{act}}^{ES} + \rho_{h,\textit{act}}^{ES} - 1) - \kappa^{ES} \right] E^{ES} \label{Ees}\\
\dot{\rho}_{b,\textit{act}}^m = & - \frac{g_m}{\nu_m Z^{QD} f^{act}} (\rho_{e,\textit{act}}^m + \rho_{h,\textit{act}}^m - 1) \frac{|E^m|^2}{\eta_m^2} \nonumber \\
		&- W_m \rho_{e,\textit{act}}^m \rho_{h,\textit{act}}^m + S^{m,cap}_{b,\textit{act}} + S^{m,rel}_{b,\textit{act}} \label{Erho}\\
\dot{\rho}_{b,\textit{inact}}^m = & - W_m \rho_{e,\textit{inact}}^m \rho_{h,\textit{inact}}^m + S^{m,cap}_{b,\textit{inact}} + S^{m,rel}_{b,\textit{inact}}\\
  \dot{w_b} = & + J - R_{loss}^W w_e w_h \nonumber \\
	      & -2N^{QD} \left(f^{\textit{act}}S^{GS,cap}_{b,\textit{act}} + f^{\textit{inact}}S^{GS,cap}_{b,\textit{inact}}\right) \nonumber \\
	      & -4N^{QD}  \left(f^{\textit{act}}S^{ES,cap}_{b,\textit{act}} + f^{\textit{inact}}S^{ES,cap}_{b,\textit{inact}}\right) \label{Ewe}
\end{align}

We consider optical injection into the ground state (see sketch of the setup in Fig.\ref{fig:setup}) and therefore the phase dynamics of the ES can be neglected.
The relative injection strength $K$ is normalized to a combined electric field amplitude $E_0$, which is defined via the the total out-coupled power
\begin{align}
2 \kappa^{GS} |E_0|^2 := \sum_{m \in \{GS,ES\}} 2\kappa^m |E^m|^2 \bigg|_{K=0} \propto P_{out}
\end{align}
This normalization ensures that the injection strength does not depend upon the distribution of the GS and ES to the total power. $K=1.0$ then corresponds to injecting as much electric field intensity into to the GS as is being coupled out from the GS and the ES. The relative frequency detuning of the injected signal from the solitary GS frequency is given by $\Delta \nu_{inj} = (\omega_{inj} - \omega_{las}^{GS}) / (2 \pi) $.

\begin{figure}
  \centering
  \includegraphics[width=0.7\columnwidth]{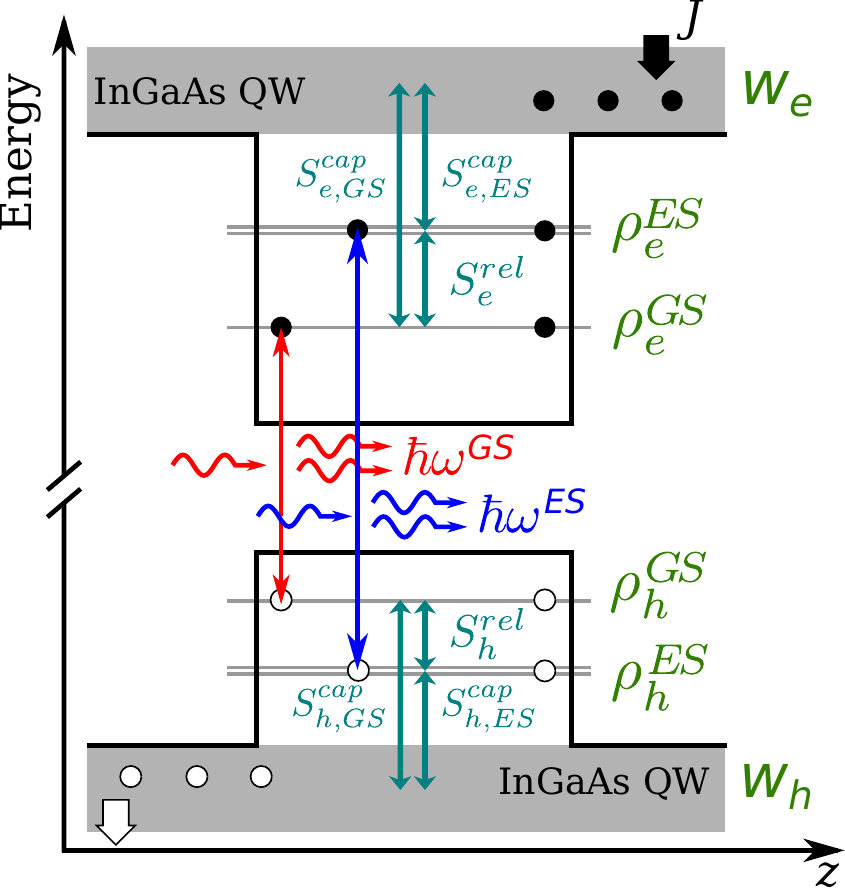}%
  \caption{Energy band structure across one QD within the self-organized layer of heterogeneously distributed QDs. Green arrows indicate carrier scattering transitions while blue and red straight arrows indicate optical transmissions. Blue and red wavy arrows sketch stimulated emission of photons. The green symbols on the right denote the dynamic charge-carrier variables described with Eq.\,(\ref{Erho})-(\ref{Ewe}).}
  \label{fig:BandStructure}
\end{figure}

The  charge-carrier variables inside the active medium and their interaction processes needed for their dynamic description are shown in Fig.\,\ref{fig:BandStructure}. There are six different carrier types: electrons and holes inside the ground state $(\rho_{e}^{GS},\rho_{h}^{GS})$, the excited state $(\rho_{e}^{ES},\rho_{h}^{ES})$ and inside the surrounding quantum-well reservoir $(w_e,w_h)$. The reservoir is pumped with an electric current density $J$. Due to the inhomogeneous broadening not all QD transitions are resonant with the optical mode and we therefore introduce a separation into resonant active carriers $\rho_{\textit{act}}$ and non-resonant carriers $\rho_{\textit{inact}}$ that do not directly interact with the light mode but indirectly via the common carrier reservoir. The resulting equations are given in Eg.\eqref{Erho}-\eqref{Ewe}. We use $b\in\{e,h\}$ and $x \in \{\textit{act},\textit{inact}\}$ to indicate the carrier type and $m \in \{GS,ES\}$ to indicate the state.
Short explanations of the parameters and the values used for numerical calculations are given in Tab.\,\ref{tab:parameters}.

The instantaneous frequency shift $ \delta \omega$ given in Eq.\,(\ref{Egs}) of the GS lasing mode is given by
\begin{align}
  \delta \omega = \delta \omega_{ES} (\rho_e^{ES} + \rho_h^{ES}) + \delta \omega_{QW}^e w_e + \delta \omega_{QW}^h w_h
  \label{eq:PAcoupling}
\end{align}
with parameters $\delta \omega_{ES}, \delta \omega_{QW}^e, \delta \omega_{QW}^h$ as given in Tab.\,\ref{tab:parameters}. We additionally include the artificial scaling parameter $\gamma$ to investigate the consequences of stronger amplitude-phase coupling (the default value is $1$). The microscopic treatment of the charge-carrier dynamics is important for an accurate modelling of the two-mode lasing characteristics of our QD laser device \cite{ROE15a}. We found the carrier scattering to only weakly depend on small variations of, e.g., the energy structure of the QD or their inhomogeneous broadening. On the other hand, the carrier-induced refractive index change and, thus, the amplitude-phase coupling depends sensitively on those parameters. While an isolated rescaling of our dynamic amplitude-phase coupling with $\gamma$ is not a rigorous microscopic treatment of the QD laser carrier dynamics, it nevertheless gives reliable results on the dynamic changes expected for QD laser structures with different amplitude-phase coupling strength.

\begin{figure}
  \includegraphics[width=\columnwidth]{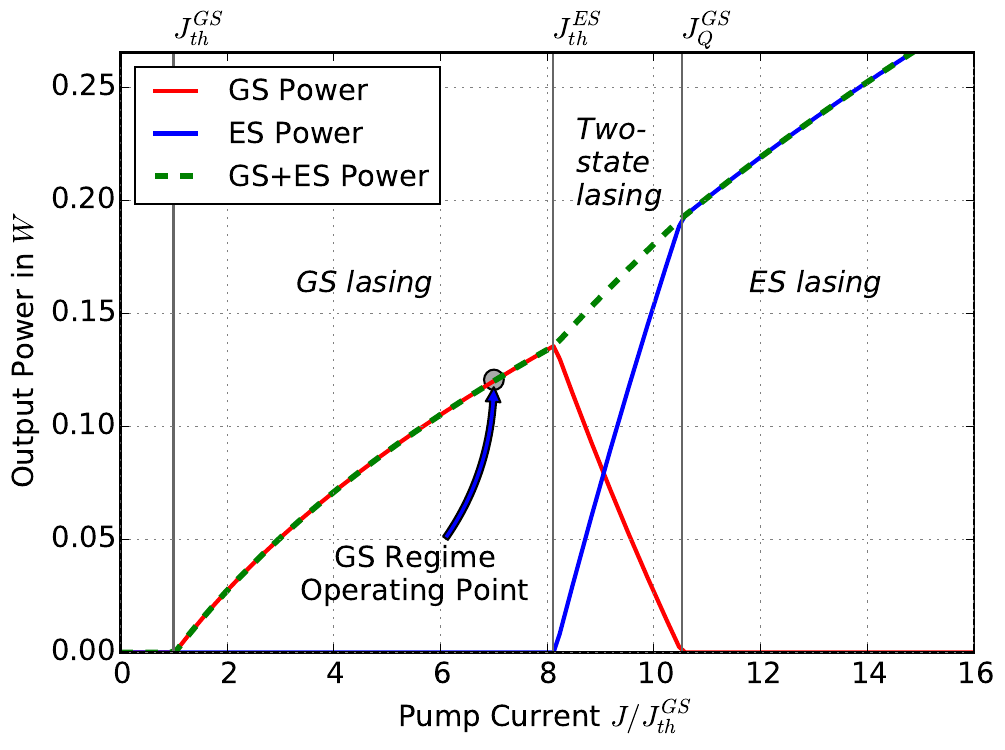}%
  \caption{Input-output characteristics of the QD-laser without optical injection. Total outcoupled power (dashed green) and GS (red) and ES (blue) outcoupled powers are plotted as a function of the pump current $J$ normalized to the threshold current of the GS ($J_{th}^{GS}$).  With increasing pump current three dynamic regimes can be distinguished. They are indicated by the vertical lines and are labeled \textit{"GS lasing"}, \textit{"two-state lasing"} and \textit{"ES lasing"}. The bended arrow shows the operating point for Figs.\,\ref{fig:2D_GSwES}-\ref{fig:2D_Bistability}.}\label{fig:LC}
\end{figure}

The scattering rates for the charge-carrier exchange processes between QD states and the quantum well reservoir are implemented via nonlinear fits to microscopically calculated rates (at $T=300K$) (see \cite{LIN14} for details) and enter the carrier equations according to Eq.\,\eqref{Scap}-\eqref{Srel}.
\begin{align}
  S_{b,x}^{m,cap} & = S_{b}^{m,in}(1 - \rho_{b,x}^m) - S_{b}^{m,cap,out}(\rho_{b,x}^m) \label{Scap}\\
  S_{b,x}^{m,rel} & = \pm \frac{1}{\nu_m} \left[ S_{b,in}^{rel} (1 - \rho_{b,x}^{GS}) \rho_{b,x}^{ES} - S_{b,out}^{rel} (1 - \rho_{b,x}^{ES}) \rho_{b,x}^{GS} \right] \label{Srel}
\end{align}

\begin{table}
  \centering
    % \begin{andptabular}[\linewidth]{l l l}{Parameters used for our numerical calculations unless otherwise noted.}
    \caption{Parameters used for our numerical calculations unless otherwise noted.}
  \begin{tabular}{l | l | l}
  \hline
  Symbol & Value & Meaning \\ \hline
   $g_{GS}$		& $230ns^{-1}$		& GS linear gain \\
   $g_{ES}$		& $460ns^{-1}$		& ES linear gain \\
   $\kappa^{GS}$	& $50ns^{-1}$		& GS optical losses \\
   $\kappa^{ES}$	& $10ns^{-1}$		& ES optical losses \\
   $Z^{QD}$		& $3.0 \times 10^7$		& number of QDs \\
   $N^{QD}$		& $10^{11}cm^{-2}$	& area density of QDs \\
   $f^{act}$		& $0.5$		& fraction of active QDs \\
   $W_{GS}$		& $0.44ns^{-1}$	& GS spont. emission rate \\
   $W_{ES}$		& $0.55ns^{-1}$	& ES spont. emission rate \\
   $R_{loss}^W$		& $0.54 \times 10^{11}cm^2ns^{-1}$	& quantum well loss rate \\
   $\eta_{GS}^2$	& $101.1 V^2cm^{-2}$	& conversion factor \\
   $\eta_{ES}^2$	& $108.5 V^2cm^{-2}$	& conversion factor \\
   $\nu_m$		& 1(GS),2(ES)		& QD state degeneracy \\
   $\gamma$		& $1$			& scaling factor \\
   $\delta \omega_{ES}$	& $125ns^{-1}$	& frequency shift coefficient \\
   $\delta \omega_{QW}^e$	&$11.3 \times 10^{-11}cm^2ns^{-1}$	& frequency shift coefficient \\
   $\delta \omega_{QW}^h$	&$5.5 \times 10^{-11}cm^2ns^{-1}$	& frequency shift coefficient
\end{tabular}
\label{tab:parameters}
% \end{andptabular}
%   \caption{Parameters used for our numerical calculations unless otherwise noted.}
% \label{tab:parameters}
\end{table}

The GS and ES electric field amplitudes are only indirectly coupled via charge-carrier scattering processes between the GS and ES QD states. These processes are slow when compared to quantum well lasers \cite{BIM08a} and thereby allow for an \textit{incomplete gain clamping} of the ES inversion \cite{MAR03c}, which enables \textit{two-state lasing}.

The simulated input-output diagram (LI-curve) of the two-state QD laser is shown in Fig.\,\ref{fig:LC}. The total out-coupled power is shown with a dashed green line while GS and ES out-coupled powers are indicated with solid red and blue lines, respectively. The ground state shows GS quenching (decreasing red line) as soon as the ES turns on. At high currents the GS is off and only ES lasing remains. We thus have three different operating regimes that we indicate by: \textit{"GS", "two-state"} and \textit{"ES" -lasing} in Fig.\,\ref{fig:LC}.  This behavior mimics experimental results \cite{MAR03a}. In a model with separate GS and ES carrier populations, it can only be reproduced, if an asymmetry in the electron and hole dynamics is included \cite{ROE15a}. Simpler two-mode laser models, which take into account only one charge-carrier type, have to include nonlinear gain terms with self- and cross-compression contributions to reproduce a similar qualitative shape of the input-output curve \cite{BRA11}. Such an approach, however, fails to describe the asynchronous refractive index dynamic of QD lasers \cite{LIN12b}, which requires multiple carrier types to be modeled. Therefore, a relatively complex model with in total 10 different carrier types is chosen.

Note that without optical injection the QD laser always emits stable light and does not exhibit bistability.
For the numerical studies of the stability an operating current within the "GS-lasing" regime close to the onset of two-state lasing is chosen (see arrow in Fig.\,\ref{fig:LC}). The regimes and the respective threshold currents are characteristic for each device and among other parameters depend upon the optical loss rates $\kappa^{GS}$ and $\kappa^{ES}$ \cite{KAP14b,ROE15a}. Note that the ES losses $\kappa^{ES}$ are chosen to be $0.2\kappa^{GS}$ to allow for ES lasing at reasonable pump currents.

\section{Injection Dynamics}

\begin{figure}
  \includegraphics[width=\columnwidth]{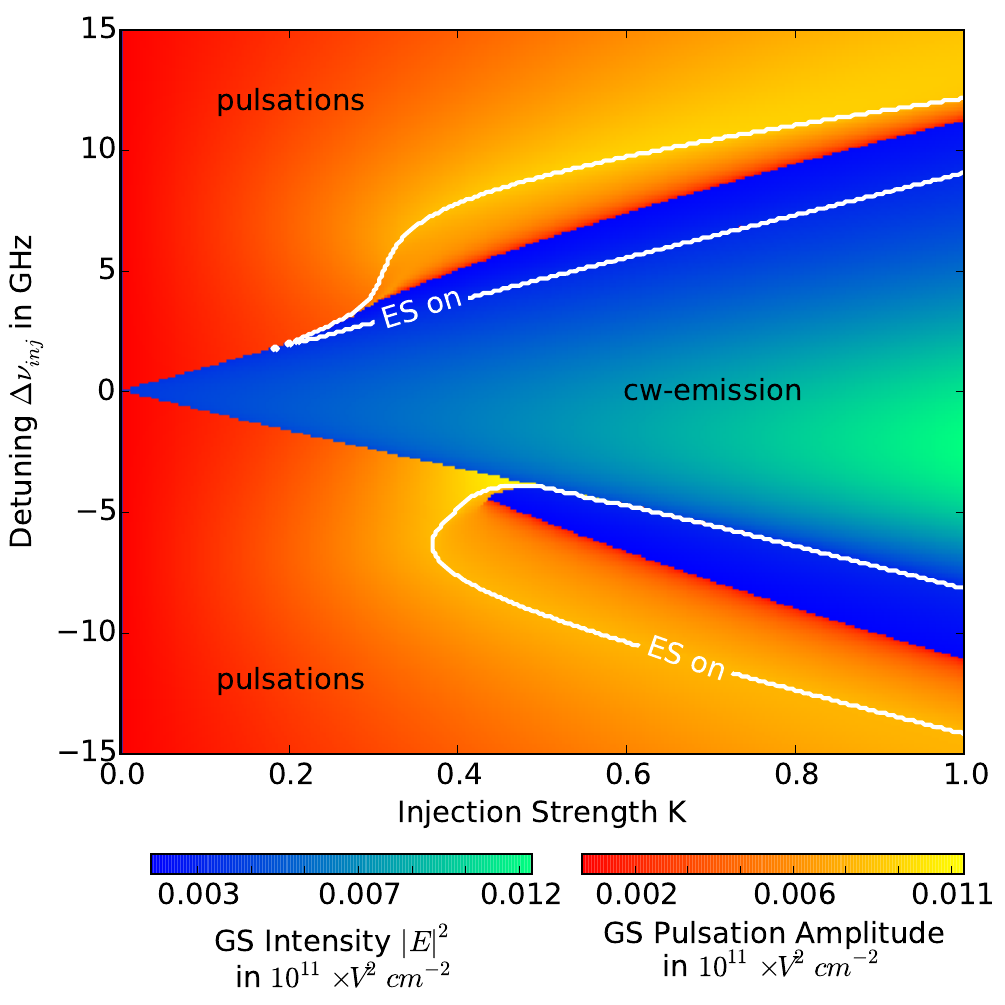}%
  \caption{Dynamics under optical injection at the GS wavelength as a function of the injection strength $K$ and the injected frequency (detuning $\Delta \nu_{inj}$ from the GS wavelength). Color coded in blue/green is the intensity of the GS within the locking region. Red/orange colors indicate the intensity of the injection induced pulsations of the GS. Solid white lines mark the region where injection induces two-state lasing. Parameters: $J\approx 7.0J_{th}^{GS}$, other parameters as given in Tab.\,\ref{tab:parameters}.}\label{fig:2D_GSwES}
\end{figure}

We first characterize the dynamics of the optically injected two-state QD laser that is operated below the ES threshold. Thus, without injection only one-color GS emisison is observed. The control parameters that are varied are the injection strength $K$ and the relative frequency detuning $\Delta \nu_{inj}$. The results are plotted in Fig.\,\ref{fig:2D_GSwES}. The blue-green color code shows the intensity of the GS within the locking region, i.e. where the GS still emits stable \textit{cw} light even with injection. Outside this locking region pulsations are observed and their pulsation amplitudes are indicated by the red/orange colors.
For $K \lesssim 0.3$, the laser unlocks in a saddle-node-infinite-period (SNIPER) bifurcation \cite{KEE81b}, where two fixed points on a periodic orbit collide and annihilate each other and only a limit cycle solution remains. As a result spiking dynamics are observed close to the bifurcation \cite{WIE05}. For $K \gtrsim 0.3$, the \textit{cw}-solution loses its stability in a Hopf bifurcation \cite{AND66}, where a stable limit cycle solution is born. Hence, undamped relaxation oscillations are found close to the bifurcation.
The signature of the bifurcations at the edge of the locking cone can be seen in the pulsation amplitude in Fig.\,\ref{fig:2D_GSwES}: The Hopf bifurcation shows a square root like increase of the pulsation amplitudes while at a SNIPER the amplitude is about constant at the bifurcation line.

Moreover, injection into the GS can induce two-state lasing even though the laser is electrically pumped below the ES threshold. For large enough detunings, the injection acts destructively on the GS and thus increases the effective losses for the GS. This, in turn, increases the carrier densities in the QD, allowing the ES to reach a sufficient inversion for lasing. We observe regions where GS and ES simultaneously emit stable \textit{cw}-lasing as well as regions where both perform pulsations. Note that due to their indirect coupling via the charge-carriers, pulsations of the GS electric field will always cause the ES electric field to also perform pulsations. The regions where ES lasing occurs are indicated with solid white lines in Fig.\,\ref{fig:2D_GSwES}. At the boundary of the two-state lasing region (solid white lines), the laser either crosses a transcritical bifurcation of fixed points (white line inside blue region) or a transcritical bifurcation of limit-cycles (white lines inside orange region). In both cases the two-state solution and the single-color solution exchange stability. These transcritical bifurcations are also found for the conventional optically injected two-mode laser by Osborne et al. \cite{OSB09,OSB09a}, however, the shape of the bifurcation lines in the ($K-\Delta \nu_{inj}$)-parameter plane differs, as those lasers already emit two modes without injection and crucial parameters like carrier lifetimes and amplitude-phase coupling are different.

In general, the locking cone in the ($K-\Delta \nu_{inj}$)-parameter plane observed for the GS emission of the optically injected QD laser that is operated at a pump current below the ES lasing threshold is similar to what is known from normal single color lasers. However, we only observe period-one pulsations outside of the locking cone and neither complex dynamics nor chaos, which are usually found in optically injected single-mode \cite{WIE05, PAU12, OSH14} and dual-mode \cite{OSH14} semiconductor lasers. We will explore the reasons for this increase of dynamic stability in the next section.

\section{Stability}
\begin{figure}
  \includegraphics[width=\columnwidth]{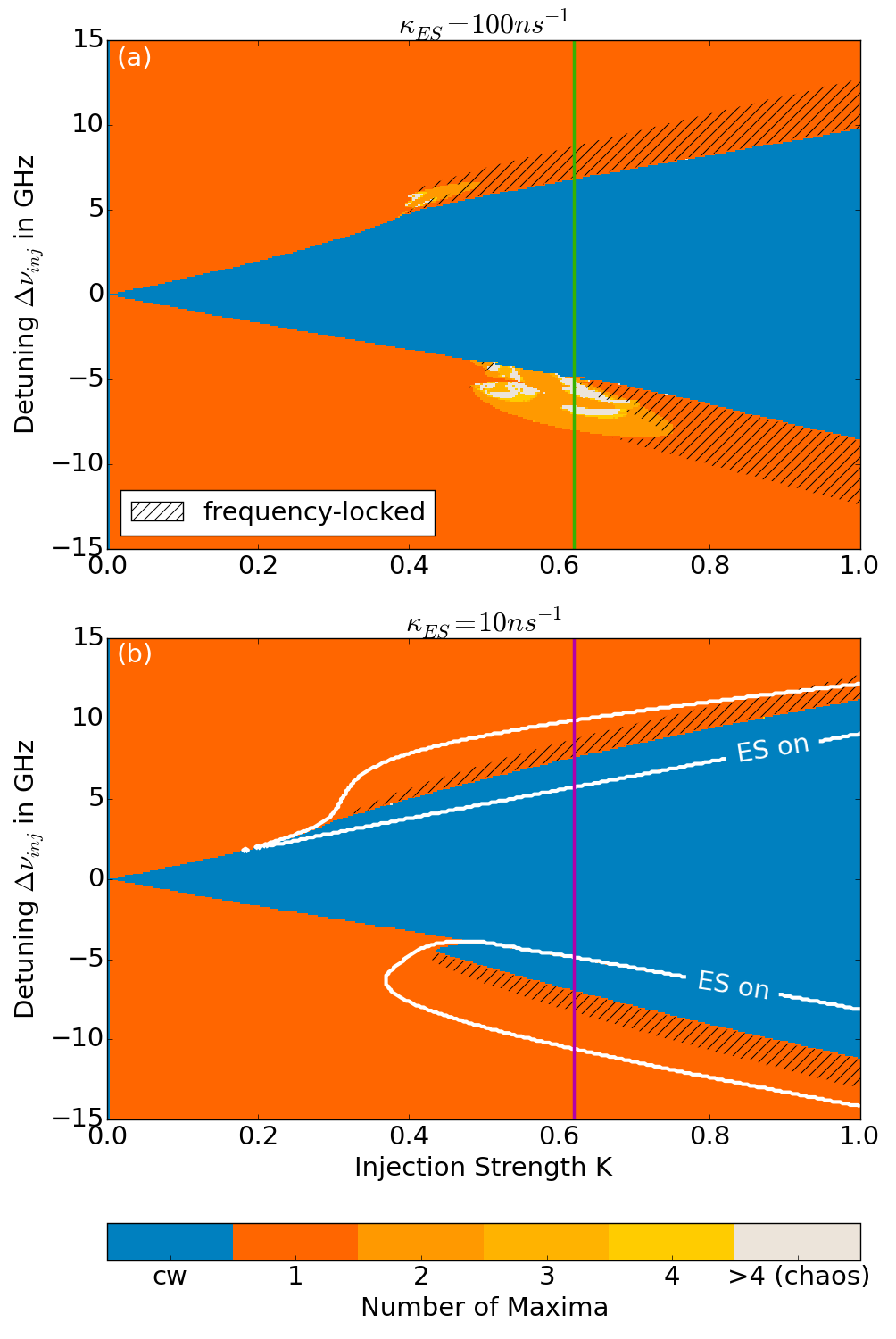}%
  \caption{Comparison between a single-color laser (a) and a two-state laser (b) under optical injection. The color code denotes the number of different maxima found in the time series of the emitted light as a function of injection strength $K$ and optical detuning $\Delta \nu_{inj}$ (blue indicates stable lasing and orange to yellow periodic pulsations with increasingly complicated pulse trains). Hatched regions indicate frequency locking. The vertical lines indicate the parameter scans presented in Fig.\,\ref{fig:DetuningLinescan}. Parameters: $J\approx 7.0J_{th}^{GS}$, other parameters as given in Tab.\,\ref{tab:parameters}.}\label{fig:DynComp}
\end{figure}

To compare the impact of two-state lasing on the dynamic stability, we compare two different QD lasers under optical injection: first, the two-state QD laser on the edge of two-state lasing as discussed in the previous section and, second, an identical QD laser with increased ES losses $\kappa_{ES}$. This way, ES emission of the QD laser is suppressed in the second case, and, thus, the typical behaviour of an optically injected single-color QD laser that has been studied before \cite{WIE05,LIN14} is recovered. The observed dynamics of this single-color QD laser are shown in Fig.\,\ref{fig:DynComp}a. Note, the color code now indicates the number of maxima found in the time series. Blue corresponds to steady state (\textit{cw})-lasing and the areas plotted in beige correspond to chaotic and high-order limit-cycle dynamics. The well known period-doubling cascade into chaos can be found at the negative detuning side in Fig.\,\ref{fig:DynComp}a. There, regions of dynamic instability e.g. higher period oscillations,
irregular oscillation and chaos, appear. The hatching indicates the regions with frequency locking (locking of the time-averaged frequency) that also occur if the laser is pulsing.

In contrast to the single-color case, the two-state QD laser shown in Fig.\,\ref{fig:DynComp}b only exhibits stable \textit{cw}-lasing (blue) and period-one oscillations (red) and lacks any regions of dynamic instabilities. For parameter values where the single-color QD laser exhibits complex dynamics, we find that the ES turns on instead for the two-state QD laser case and stable two-wavelength emission is found. Consequently the locking cone becomes wider.  This is somewhat surprising and counter-intuitive. The added degree of freedom in the two-state laser leads to less complicated dynamics. A similar trend was also observed for conventional optically injected Fabry P\'{e}rot lasers that support two longitudinal modes. Starting from a much more unstable behaviour of the one-color laser the ability for two-mode emission also lead to suppressed chaotic regions and favoured two-mode emission \cite{OSH14}.

\begin{figure}
  \includegraphics[width=\columnwidth]{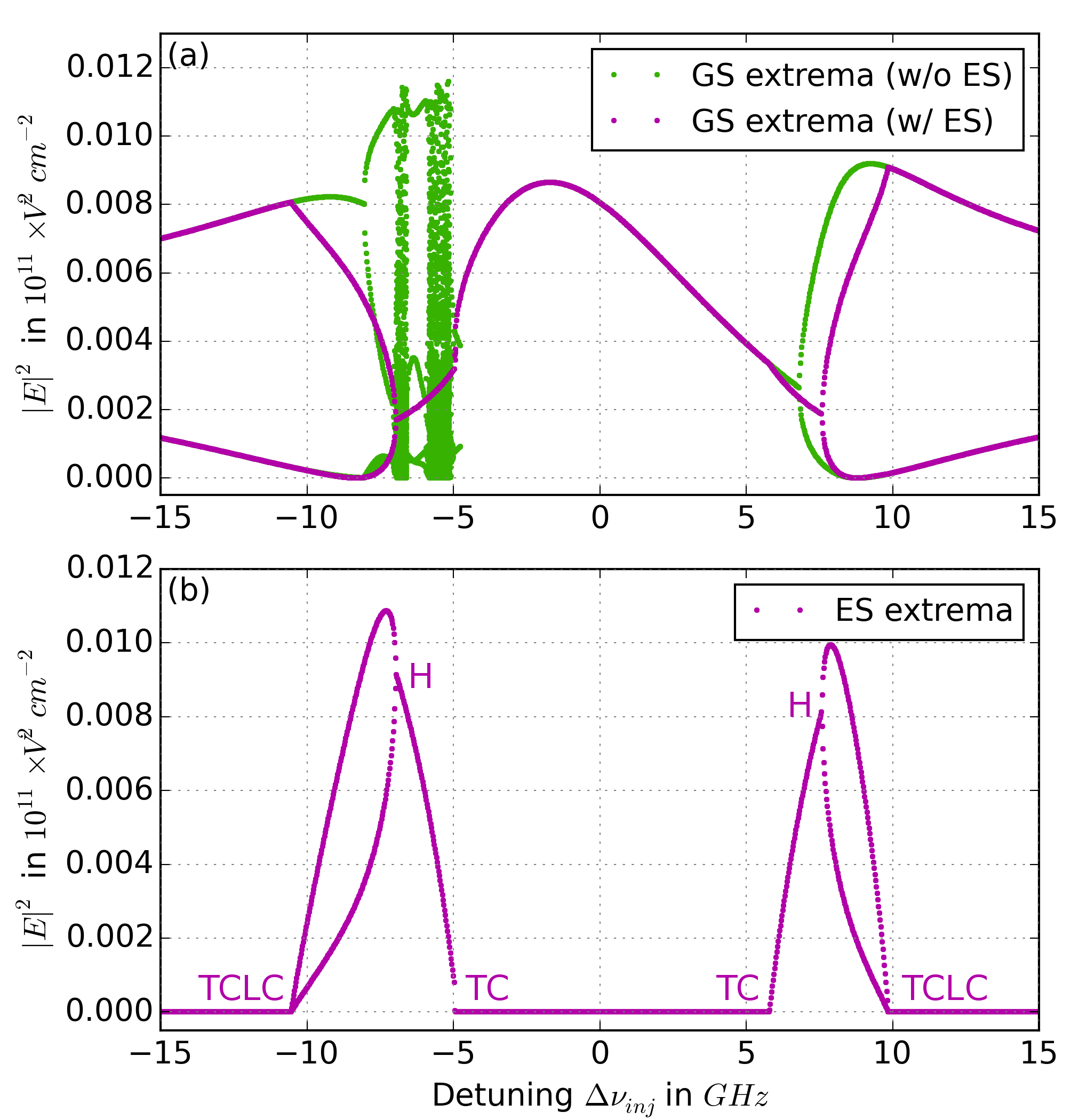}%
  \caption{\label{fig:DetuningLinescan}
    Bifurcation diagram: Linescan along increasing detuning (indicated by the vertical line in Fig.\,\ref{fig:DynComp}). Violet and green dots show extrema found in the time series of the two-color and single-color laser, respectively. (a) Extrema found in the GS intensity of both lasers. (b) Extrema found in the ES intensity of the two-state laser. 'H' indicates a Hopf-bifurcation, 'TC' a transcritical bifurcation and 'TCLC' a transcritical bifurcation of a limit cycle. An increase of the dynamic stability is found if two-state lasing is possible. Parameters: $K = 0.62$, $J\approx 7.0J_{th}^{GS}$, other parameters as given in Tab.\,\ref{tab:parameters}.}
\end{figure}

To elaborate the stabilizing effect of ES lasing further, Fig.\,\ref{fig:DetuningLinescan}a shows line scans of the dynamic response of the GS emission to optical injection (dots correspond to extrema found in the time series) for the case of the two-state laser (violet) and the single-color laser (green). The scans are performed for varying $\Delta \nu_{inj}$ along the vertical green and purple lines in Fig.\,\ref{fig:DynComp}. Figure~\ref{fig:DetuningLinescan}b shows the corresponding dynamics of the ES emission from the two-state laser. For detunings between $-7\text{GHz}$ and $-5\text{GHz}$ the single color laser exhibits chaotic pulsing whereas the two-state laser shows stable  \textit{cw}-emission. Thus, the locking range of the the two-state laser, extending up to $\Delta \nu_{inj} \approx -7 \text{GHz}$, is larger than the locking range the single-color laser, which only extends to $\Delta \nu_{inj} \approx -5\text{GHz}$. Although no chaos is observed on the positive detuning side in the single-color laser, the two-state laser there also remains locked up to greater detunings than the single-color laser ($\Delta \nu_{inj} \approx 7.6\text{GHz}$ vs. $\Delta \nu_{inj} \approx 6.8\text{GHz}$).

Inspecting the two-state laser ES dynamics in Fig.\,\ref{fig:DetuningLinescan}b illustrates how ES lasing corresponds to the stabilization of the GS. We observe stable ES \textit{cw}-emission between $5\text{GHz} \lesssim |\Delta \nu_{inj}| \lesssim 7\text{GHz}$, period-one pulsations between $7\text{GHz} \lesssim |\Delta \nu_{inj}| \lesssim 10\text{GHz}$ and no ES lasing otherwise. The parameter values where the ES turns on with $cw$ emission are the signatures of the transcritical bifurcation (denoted with 'TC') while it turns on with periodic oscillations in a transcritical bifurcation of limit cycles (denoted with 'TCLC'). The stable two-state emission that is born at TC loses its stability in a Hopf bifurcation (denoted with 'H'). The bifurcations of the single-color laser (green dots in Fig.\ref{fig:DetuningLinescan}a) involve a period doubling route to chaos at the negative detuning side as e.g. described in \cite{PAU12}, while the single mode lasing loses stability in a Hopf bifurcation for positive detunings. For increasing negative detunings we find that the turn on of the ES (at TC)  coincides with the emergence of chaos in the single-color QD laser case (seen as the densely packed green regions in Fig.\,\ref{fig:DetuningLinescan}a), while the Hopf bifurcation of the ES ($\Delta \nu_{inj} \approx -7\text{GHz}$) marks the point where the single-color laser restabilizes to periodic oscillations.

So far our numeric results did not show bistability between the different sweep directions. However, an optically induced hysteresis has been observed experimentally in a two-state QD laser \cite{TYK16a}. In the following, we therefore look for bistability in our two-state QD laser model.

\section{Bistability}

\begin{figure}
  \includegraphics[width=\columnwidth]{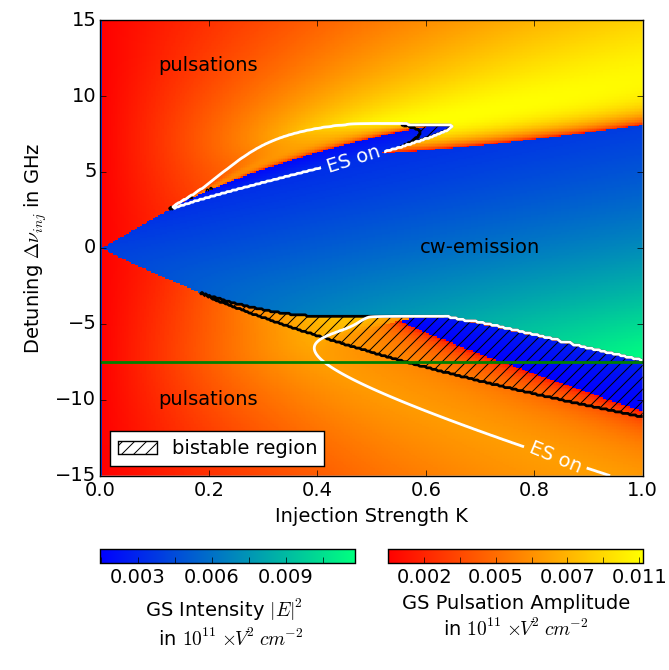}%
  \caption{Impact of amplitude-phase coupling: Same as Fig.\,\ref{fig:2D_GSwES} but with an increased $\gamma=4$ to yield a higher effective $\alpha$-factor of $2.08$.The hatched area indicates bistability between ES on and ES off. The injection strength $K$ is swept and the up direction is shown. Along the horizontal line, bifurcation diagrams are evaluated in Fig.\,\ref{fig:1D_Bistability}a. Parameters: $J\approx 7.0J_{th}^{GS}$, other parameters as given in Tab.\,\ref{tab:parameters}.
}\label{fig:2D_Bistability}
\end{figure}

Studies of simpler semiconductor-laser models have linked the emergence of bistability under injection to a sufficiently strong amplitude-phase coupling ($\alpha$-factor) \cite{GAV97}. The strength of the amplitude-phase coupling in QD devices sensitively depends on the energy level spacings between the GS, the ES and the quantum well, which are determined by the material system and physical size of the QDs \cite{LIN13,GIO07}. The scaling factor $\gamma$ introduced in Eq.\,\eqref{Egs} allows us to phenomenologically model those effects, while keeping the microscopically derived dependencies on the carrier dynamics. Our numerical investigations then show, that a lower pump current or an increase of the amplitude-phase coupling leads to the emergence of bistability between ES and GS lasing.

Calculating an effective $\alpha$-factor allows for comparisons of our microscopically based amplitude-phase coupling strength with more conventional models, that use the constant $\alpha$-factor approximation. Generally, the $\alpha$-factor is defined via changes of the optical susceptibility with respect to the total carrier number. This approach however, fails for QD lasers \cite{LIN12b} and thus the effective $\alpha$-factor is defined via the response of the complex gain $g$ of the laser to an injected signal \cite{LIU01b}:
\begin{align}
 \alpha_{\text{\text{eff}}} = -\frac{\frac{\partial}{\partial K} \operatorname{Im} (g)}{\frac{\partial}{\partial K} \operatorname{Re} (g)}\Bigg|_{K=0} = -\frac{\frac{\partial}{\partial K} \gamma \delta \omega}{\frac{\partial}{\partial K} g_{GS}(\rho_{e,act}^{GS} + \rho_{h,act}^{GS} -1) } \Bigg|_{K=0}\label{eq:alpha}
\end{align}
$\gamma = 1$, which was used so far, is the microscopically calculate value and yields an effective $\alpha \approx 0.5$. In the following, we investigate the dynamics found for a higher effective $\alpha$-factor $\alpha_{eff} \approx 2.0$, corresponding to the higher end of commonly measured $\alpha$-factors in QD lasers \cite{GER08, LIN13a}. We implement this by choosing $\gamma=4$.

Figure \ref{fig:2D_Bistability} shows the dynamics with an amplitude-phase coupling increased by a factor of four. As a result the instantaneous lasing frequency responds stronger to optical injection and a large region of bistability appears at the lower edge of the locking cone ($\Delta \nu_{inj} \lesssim -3.5GHz$ and $K \gtrsim 0.2$, denoted by the black hatching, which will be discussed in further detail.

\begin{figure}
  \includegraphics[width=\columnwidth]{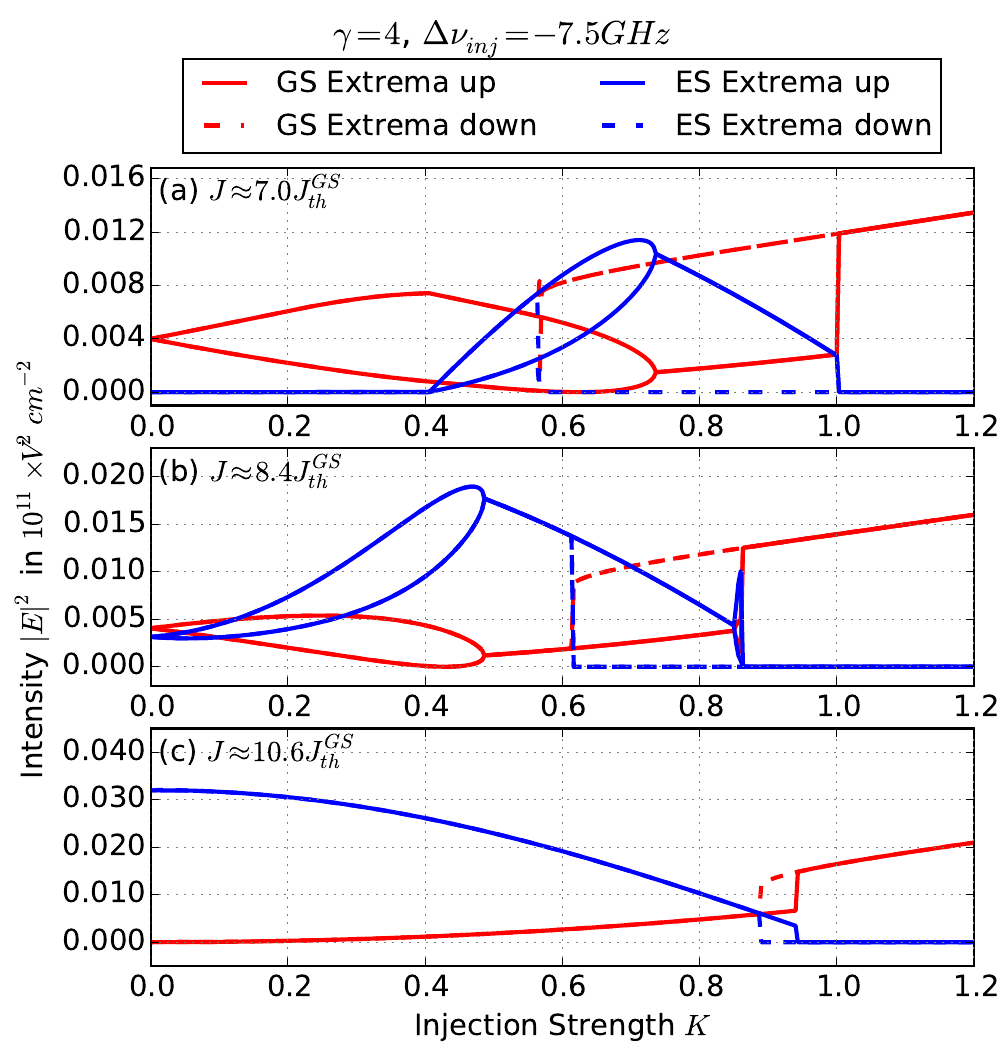}%
  \caption{\label{fig:1D_Bistability}
   Hysteresis within the region of bistability for different pump currents: (a) \textit{GS lasing} $J\approx 7.0J_{th}^{GS}$, (b) \textit{two-state lasing} $J\approx 8.4J_{th}^{GS}$ and (c) \textit{ES lasing} regime $J\approx 10.6J_{th}^{GS}$. The solid lines show  up-sweeps, the dashed lines down-sweeps. The size of the bistable region in the injection dimension is denoted with $\Delta K$.}
\end{figure}

Within the large region of bistability the device shows hysteresis. The laser remains on the \textit{cw} GS-only solution when decreasing the injection strength. Contrary, the lasers remains much longer in the two-state lasing regime before the ES finally turns off, when performing an up-sweep in $K$ . This is emphasized by a linescan shown in Fig.\,\ref{fig:1D_Bistability}a (taken along the green horizontal line indicated in Fig.\,\ref{fig:2D_Bistability} at $\Delta \nu_{inj}=-7.5GHz$). There, solid lines indicate the up-sweep and dashed lines indicate the down-sweep. The hysteresis loop can clearly be seen between $K=0.56$ and $K=1.00$ in Fig\,\ref{fig:1D_Bistability}a.

By changing the pump current we can navigate between the different operating regimes of the solitary QD laser as was shown in Fig.\,\ref{fig:LC}. So far the laser was operated in the \textit{GS lasing} regime. Figure~\ref{fig:1D_Bistability}b and c now show that the same kind of bistability also occurs in the \textit{two-state} (between $K=0.62$ and $K=0.87$) and \textit{ES lasing} regime (between $K=0.89$ and $K=0.95$). The bistability in the two-state regime Figure~\ref{fig:1D_Bistability}b is very similar to the one reported in \cite{OSB09a} and the bistability in the ES regime reproduces the behavior that was observed in \cite{TYK16a}. Hence two-state QD lasers combine a variety of different bistability scenarios of a similar mechanism, among which can be chosen be selecting an appropriate operating current.

\begin{figure}
  \includegraphics[width=\columnwidth]{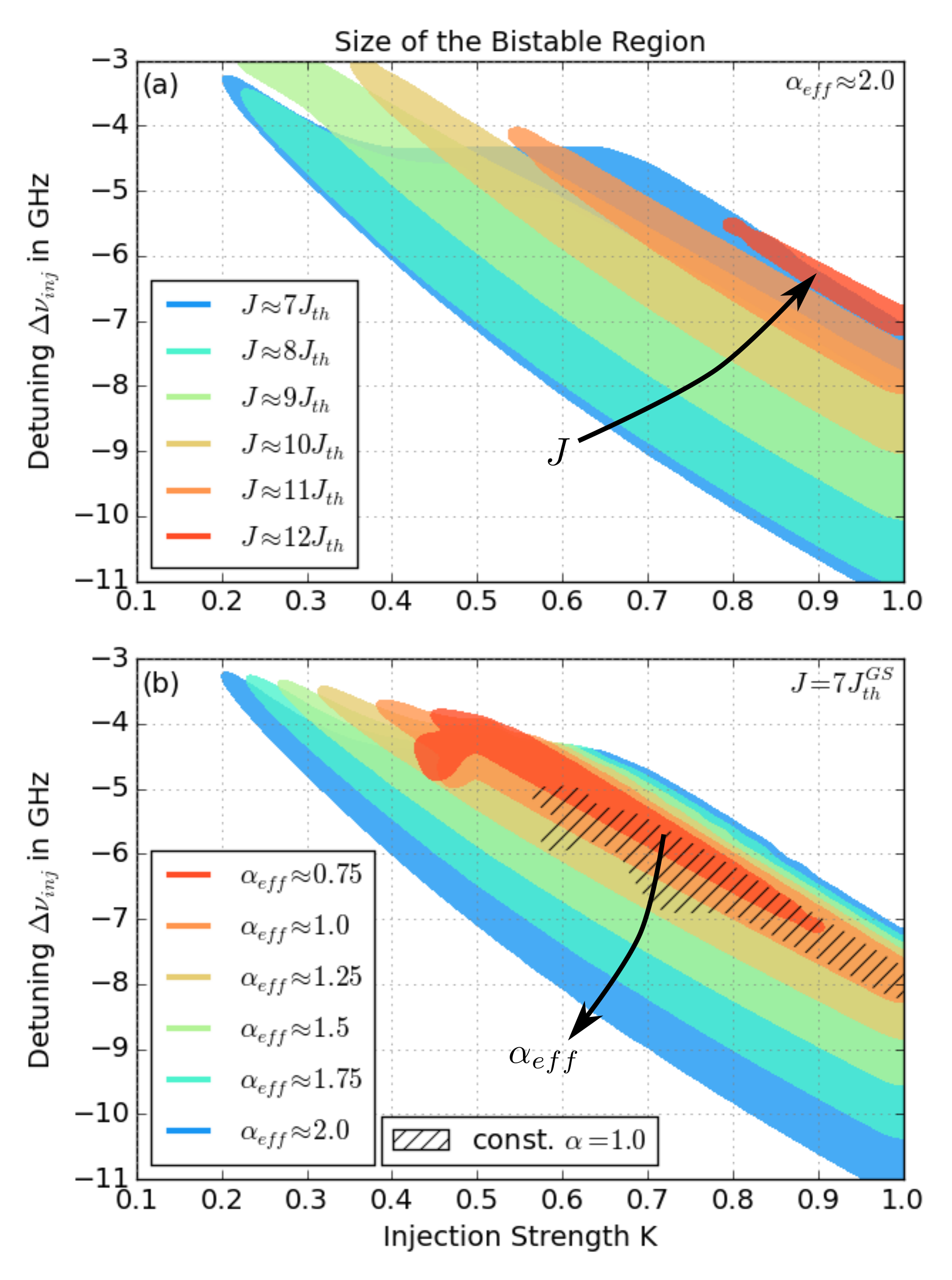}%
  \caption{\label{fig:hystsize}
    Parameter dependence of the size of the region of bistability. The colored regions in (a) indicate the size of the bistable region for different pump currents $J$ for $\gamma=4$ ($\alpha_{eff} \approx 2.0$) and in (b) the colored regions indicate different effective $\alpha$-factors for $J=7J_{th}^{GS}$. The hatched region indicates bistability with a constant $\alpha$-factor approximation. For $\alpha_{eff} \lesssim 0.75$ and $J\gtrsim12J_{th}^{GS}$ no bistability is found.}
\end{figure}

Nevertheless, it needs to be noted that the size of the bistable region decreases with pump current $J$. This is presented in Fig.\,\ref{fig:hystsize}a, showing the bistable region for different pump currents in the $K,\Delta \nu_{inj}$ plane for $\gamma=4$ ($\alpha_{eff} \approx 2.0$). For the lower pump currents $J=7J_{th}^{GS}$ and $J=8J_{th}^{GS}$ the laser is in the \textit{GS lasing} regime and exhibits the largest regions of bistability. For $J=9J_{th}^{GS}$ and $J=10J_{th}^{GS}$ the laser is in the \textit{two-state lasing} regime where the bistable regions are considerably smaller and for $J=11J_{th}^{GS}$ and $J=12J_{th}^{GS}$ the laser is in the \textit{ES lasing} regime, where the bistable regions are the smallest. Above $J=12J_{th}^{GS}$ no bistability is found. The bistability between GS and ES is therefore predominantly found in the \textit{GS lasing} regime.

On the other hand, Fig.\,\ref{fig:hystsize}b plots the size of the bistable region for different amplitude-phase coupling strengths (effective $\alpha$-factor) for $J=7J_{th}^{GS}$ (\textit{GS lasing} regime). Here, the size of the bistable region monotonously increases with the strength of amplitude-phase coupling. Below $\alpha_{eff} \approx 0.75$ ($\gamma = 1.5$) no bistability is found. Additionally, the hatching indicates a region of bistability that was found using a constant $\alpha$-factor approximation instead of the carrier dependent amplitude-phase coupling. It is similar in shape but shifted towards stronger injection and detuning when compared to the $\alpha_{eff} \approx 1.0$ region of bistability. This contrasts the results of \cite{TYK16a}, where bistability could not be achieved with a constant $\alpha$-factor approximation.

In conclusion, a strong parameter sensitivity of the existence and size of the bistable region on the amplitude-phase coupling strength and the pump current is found. To achieve bistability in two-state QD lasers, one must therefore engineer their physical sizes such that a sufficient amplitude-phase coupling exists and operate them at a pump current below the onset of solitary \textit{two-state lasing}.

\section{All-Optical Switching}

\begin{figure}
  \includegraphics[width=\columnwidth]{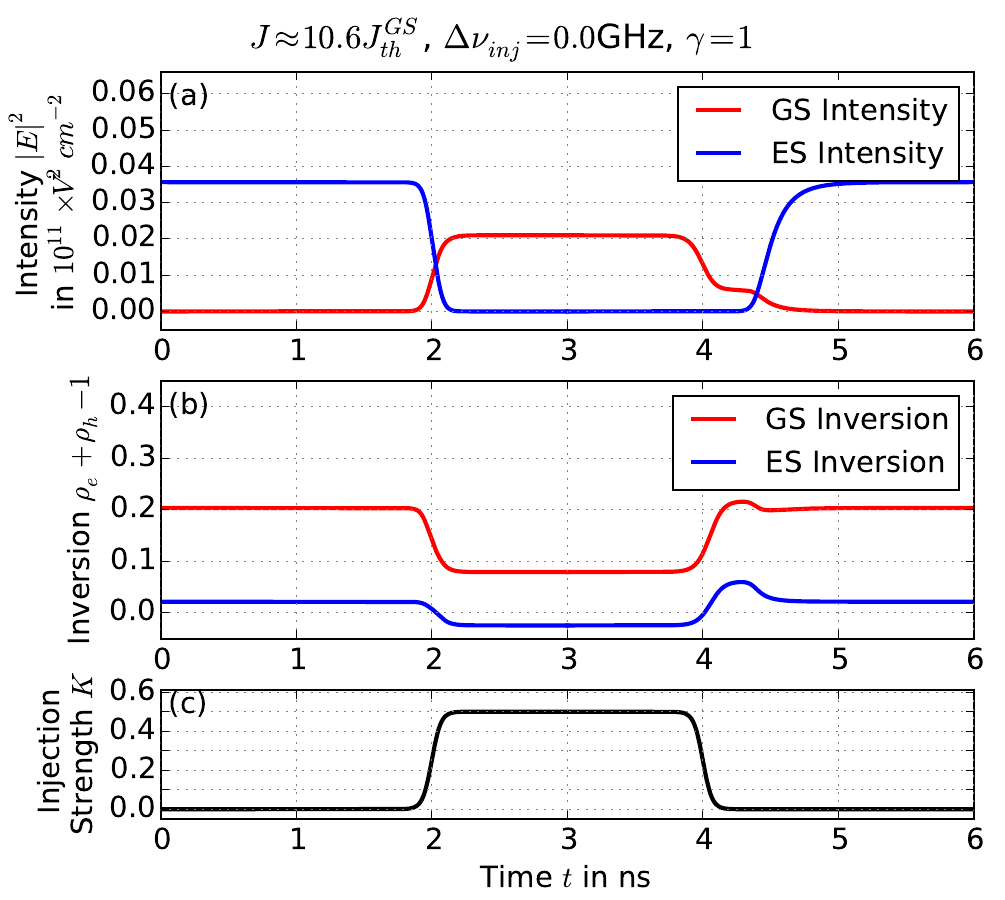}%
  \caption{\label{fig:switching}
  Switching induced by continuous injection within the ES lasing regime: (a) shows the GS (red) and ES (blue) intensity and (b) the corresponding carrier inversions. The injection strength  is increased to $K=0.5$ within the time interval indicated in (c). The switching between ES and GS occurs on a timescale of about $300ps$ and between GS and ES on a timescale of about $900ps$. The GS to ES transition is governed by strongly damped relaxation oscillations.}
\end{figure}

In this section we want to address possibilities to exploit the observed dynamics of QD lasers for all-optical switching applications. For two closely spaced longitudinal modes of a conventional laser different setups have been studied and experimentally realized \cite{OSB09a,OSB12}. Recently, also the switching between GS and ES emission of a QD laser  was experimentally demonstrated via injection into the GS \cite{TYK14}. Our model nicely reproduces these results.
A time series of the switching is shown in Fig.\,\ref{fig:switching} where the laser is operated in the \textit{ES lasing} regime and driven into \textit{GS lasing} by optical injection. The injection is smoothly switched on at $t=2$ns and off at $t=4$ns with a switching time of $\approx 100$ps. Fig\,\ref{fig:switching}a shows the electric field intensities. The timescales for the GS switch-on of about $300$ps and the GS switch-off of about $900$ps compare very well to findings in Ref.\,\cite{TYK14}.

The time series of the population inversion shown in Fig.\,\ref{fig:switching}b gives insights into the switching mechanism. Without the injection, the ground state is quenched, while the excited state is lasing and the excited state inversion is gain clamped. Injection into the ground state lowers its effective losses, thus enabling GS lasing and decreasing the GS population. As a direct consequence, the excited state inversion, which is coupled to the ground state inversion via charge-carrier scattering processes, decreases as well and falls below the lasing threshold, switching it off. As soon as the injection is turned off, the system returns to its initial state of \textit{ES lasing} and GS quenching.

\begin{figure}
  \includegraphics[width=\columnwidth]{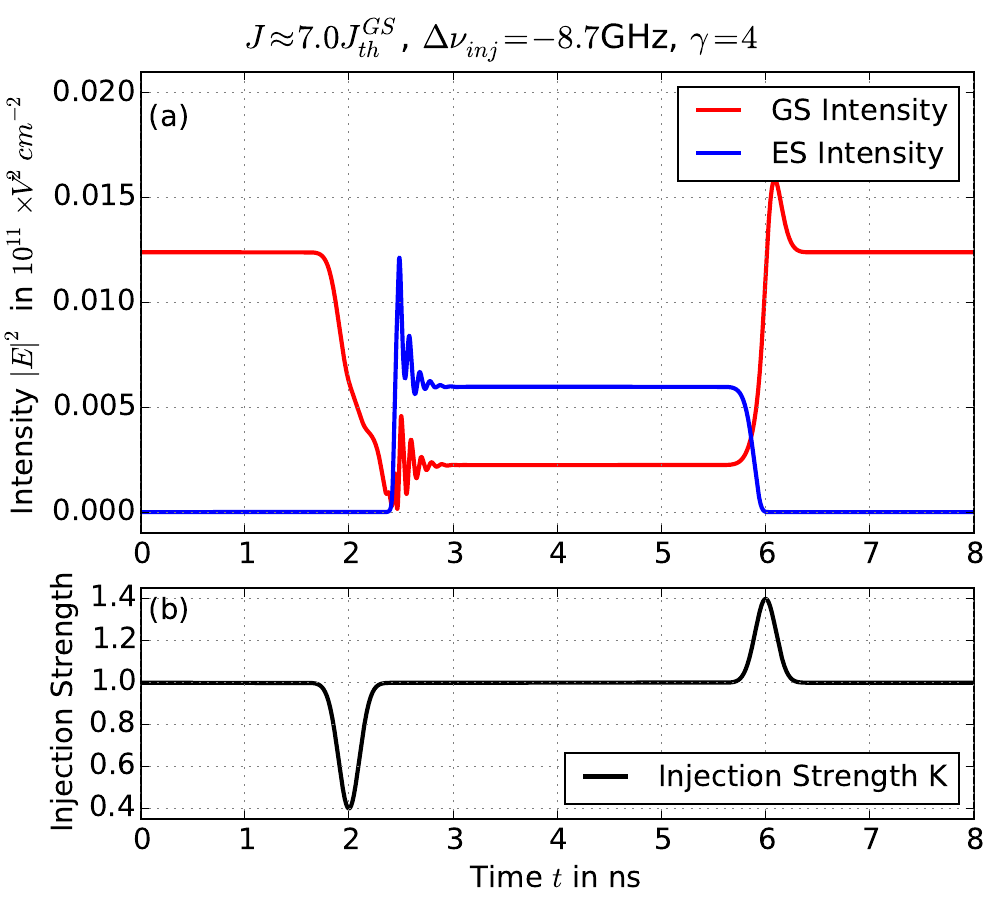}%
  \caption{\label{fig:switch}
  Switching induced by short injection pulses while the injected QD laser is operated in the bistability region: (a) ES and GS intensities (blue and red lines) during switching via short pulses in the injection strength (black line in (b)). Switching the ES on is slowed down by relaxation oscillations but occurs in less then $1$ns while switching the ES off is faster on a timescale of about $400$ps.}
\end{figure}

Furthermore, Fig.\,\ref{fig:switch} presents a different method of switching between the two states using the bistability discussed in the previous section. Now, the laser with an increased amplitude-phase coupling ($\gamma = 4$) is operated in the \textit{GS lasing} regime ($J\approx 7.0_{th}^{GS}$) and injected such that it is in a bistable region ($K = 1.0$, $\Delta \nu_{inj}=-8.7GHz$). Short pulses that increase or decrease the injection strength can then be used to switch between the two bistable solutions. Fig.\,\ref{fig:switch}b shows the pulses while Fig.\,\ref{fig:switch}a shows time traces of the GS and ES electric field intensities. Switching from \textit{GS lasing} to \textit{two-state lasing} occurs in less then $1$ns and is slowed down by relaxation oscillations while switching back does not exhibit relaxation oscillations and is considerably faster on a timescale of about $400$ps. This switching mechanism and operation is similar to what has been reported in \cite{OSB09a}. However, the strong damping of the relaxation oscillations in QD lasers compared to quantum well lasers \cite{LUE11a} enables much faster switching times.

\section{Conclusion}

Using a microscopically motivated rate equation model, we have numerically investigated the dynamic response of a two-state QD laser to optical injection. We have compared QD lasers with and without the ability to perform  two-state emission and found that the possibility of two wavelength emission drastically improves the dynamic stability when subjected to optical injection. Our results show that the added degree of freedom in two-state QD lasers leads to a high level of stability and suppresses complex dynamics.
Moreover, our model reproduces an experimentally observed bistability in optically injected QD lasers and allows for numerical investigations of the parameter dependence. We have found strong dependencies upon the pump current and the amplitude-phase coupling. The size of the bistability increases with the amplitude-phase coupling and first increases and then decreases again with the pump current. Both for too small amplitude-phase couplings and too high pump currents, the bistability vanishes.
Lastly, we have demonstrated two techniques how injection can be used for all-optical switching between different operating regimes. The first technique requires a constant change in injection strength to achieve switching and excellently matches experimental results, while the second technique exploits the bistability and only requires short trigger pulses to achieve switching. This setup could be used to implement all-optical buffers for use in future all-optical computing applications.

\section{Acknowledgment}

This work was supported by DFG within the frameworks of SFB787 and SFB910. K. Lüdge acknowledges the
Humboldt Foundation for their support in the framework of a Feodor Lynen Scholarship for Experienced Researchers.

\section*{Appendix}

\begin{table}
  \centering
  \caption{Fit parameters for charge-carrier scattering processes, extracted from microscopic calculations for 64(35)meV GS confinement energy and 50(20)meV GS-ES separation for electrons(holes) at $T=300$K.}
  \begin{tabular}{lccccc}
  % \begin{andptabular}[\linewidth]{lccccc}{\label{tab:fit}Fit parameters for charge-carrier scattering processes, extracted from microscopic calculations for 64(35)meV GS confinement energy and 50(20)meV GS-ES separation for electrons(holes) at $T=300$K.}
    \hline
    \multicolumn{1}{r}{b:}&\multicolumn{2}{c}{electrons}&&\multicolumn{2}{c}{holes}\\\cline{2-3}\cline{5-6}
    \multicolumn{1}{r}{m:}&GS&ES&&GS&ES\\\hline
    A ($10^{-11}$cm$^2$ns$^{-1}$)~~~~~~~~~~~~~~&18.5&48.3&&10.5&21.4\\
    B ($10^{11}$cm$^{-2}$)&1.9&0.48&&5.3&1.8\\
    C (ns$^{-1}$)&\multicolumn{2}{c}{1014}&&\multicolumn{2}{c}{2272}\\
    D ($10^{11}$cm$^{-2}$)&\multicolumn{2}{c}{1.4}&&\multicolumn{2}{c}{2.3}\\
    \hline
  % \end{andptabular}
  \end{tabular}
  \label{tab:fit}
\end{table}

For numerical efficiency, the nonlinear charge-carrier in-scattering rates are implemented via Eq.\,\eqref{eq:Scap} and Eq.\,\eqref{eq:Srel}, where the parameters $A,B,C$ and $D$ are fitted to the fully microscopically calculated rates \cite{MAJ11,LIN14}. The fit parameters are presented in Tab\,\ref{tab:fit}.
\begin{align}
 S_{b}^{m,cap,in}(w_b) &= \frac{A_{m,b} w_b^2}{B_{m,b} + w_b} \label{eq:Scap}\\
 S_{b}^{rel,in}(w_b) &= \frac{C_{b} w_b}{D_{b} + w_b} \label{eq:Srel}
\end{align}
The corresponding out-scattering rates are then calculated via the detailed balance conditions \cite{LUE11a}
\begin{align}
 S_{b}^{m,cap,out} &= S_{b}^{m,cap,in} \exp \left(-\frac{E_{F,b}^{eq} - \epsilon_{m,b}}{k_b T}\right) \\
 S_{b}^{rel,out} &= S_{b}^{rel,in} \exp \left(-\frac{\epsilon_{ES,b} - \epsilon_{GS,b}}{k_b T}\right)
\end{align}
where $E_{F,b}^{eq}$ is the quasi-Fermi level of the respective quantum well band, $\epsilon_{m,b}$ the confinement energy relative to the QW band edge of the corresponding QD state, $k_B$ the Boltzmann constant and $T$ the charge-carrier temperature. The quasi-Fermi levels are dynamically calculated via
\begin{align}
 E_{F,b}^{eq} = E^{QW}_{b,0} + k_b T \ln \left[ \exp \left( \frac{w_b}{\mathcal{D}_b k_b T} \right) - 1 \right]
\end{align}
where $E^{QW}_{b,0}$ is the QW band edge and $\mathcal{D}_b$ the density of states. The in and out-scattering rates then enter Eq.\,\eqref{Scap} and Eq.\,\eqref{Srel} and thereby determine the internal QD charge-carrier dynamics.

%%% Use the following two code lines if you wish to generate your bibliography with BibTeX;
%%% please replace first the string "demo" below with the name(s) of
%%% the BibTeX data base(s) you want to use.
%%% The resulting bibliography-output (the contents of the .bbl file)
%%% must be pasted into this file before submission.
%%%
 %\bibliographystyle{andp2012}
 %\bibliography{ref}
\providecommand{\WileyBibTextsc}{}
\let\textsc\WileyBibTextsc
\providecommand{\othercit}{}
\providecommand{\jr}[1]{#1}
\providecommand{\etal}{~et~al.}

\end{document}